\documentclass[twocolumn, showpacs, amsmath,amssymb]{revtex4} 
\usepackage{epsfig, amssymb}
\begin{document}

\title{An efficient algorithm to recognize local Clifford equivalence of graph states}

\author{Maarten Van den Nest}
\email{maarten.vandennest@esat.kuleuven.ac.be}
\author{Jeroen Dehaene}
\author{Bart De Moor}
 \affiliation{Katholieke Universiteit Leuven, ESAT-SCD, Belgium.}
\date{\today}

\begin{abstract}
In [Phys. Rev. A  69, 022316 (2004)] we presented a description of the action of local Clifford
operations on graph states in terms of a graph transformation rule, known in graph theory as
\emph{local complementation}. It was shown that two graph states are equivalent under the local
Clifford group if and only if there exists a sequence of local complementations which relates their
associated graphs. In this short note we report the existence of a polynomial time algorithm,
published in [Combinatorica 11 (4), 315 (1991)], which  decides whether two given graphs are
related by a sequence of local complementations. Hence an efficient algorithm to detect local
Clifford equivalence of graph states is obtained.

\end{abstract}

 \maketitle

Graph states have been studied extensively and have been employed in a number of applications in
quantum information theory and quantum computing (see e.g. \cite{Gott, graphbriegel, 1wayQC,
entgraphstate}. This is mainly due to the fact that these states can be described in a relatively
transparent way, while they maintain a sufficiently rich structure. In this note we consider the
problem of recognizing local Clifford (LC) equivalence between graph states. This issue is of
natural importance in multipartite entanglement theory \cite{entgraphstate, localcliffgraph} and in
the development of the one-way quantum computer, which is a universal measurement-based model of
quantum computation \cite{1wayQC}. In the following, we present an efficient algorithm which
recognizes whether two given graph states are LC-equivalent. At the heart of this algorithm lies an
earlier result \cite{localcliffgraph} of ours, which is a translation of the action of local
Clifford operations on graph states in terms of a graph transformation rule known in graph theory
as \emph{local complementation}. It was shown that two graph states are equivalent under the local
Clifford group if and only if there exists a sequence of local complementations which relates their
associated graphs. As it turns out, local graph complementation is well known in graph theory (see
e.g. \cite{Bouchet} and references within). What is more, in ref. \cite{alg_bouchet} a polynomial
time algorithm is derived which detects whether two given graphs are related by a sequence of local
complementations. This yields for our purposes an efficient algorithm which recognizes
LC-equivalence of graph states. We repeat this algorithm below. Note that the present result
immediately yields an efficient algorithm to recognize LC-equivalence of \emph{all} stabilizer
states (and not just the subclass of graph states). Indeed, it is well known that any stabilizer
states is LC-equivalent to a graph state \cite{stabgraphcode, graph_grassl}. Moreover, if a
particular stabilizer state is given then an LC-equivalent graph state can be found in polynomial
time, as the typical existing algorithms used to produce this graph state essentially use pivoting
methods, which can be implemented efficiently.

Before presenting the algorithm, we state some definitions and introduce some notations.
\emph{Graph states}
are special cases of stabilizer states. They are
defined as follows: let $G$ be a simple graph on $n$ vertices with adjacency matrix $\theta$
\footnote{A simple graph $G$ has no loops or multiple edges. Therefore, it can be described by a
$n\times n$ symmetric matrix $\theta$ where $\theta_{ij}$ is equal to 1 whenever there is an edge
between vertices $i$ and $j$ and zero otherwise. As $G$ has no loops, $\theta_{ii}=0$ for every
$i=1, \dots, n$} and define $n$ commuting correlation operators $K_j$ ($j=1,\dots, n$), which act
on $\Bbb{C}_2^{\otimes n}$, by
\[K_j := \sigma^{(j)}_x \prod_{k=1}^n \left(\sigma_z^{(k)}\right)^{\theta_{kj}}.\]  Here $\sigma^{(i)}_x, \sigma^{(i)}_y, \sigma^{(i)}_z$ are the Pauli
matrices which act on the $i$th copy of $\Bbb{C}_2$. The graph state $|G\rangle$ is the unique
eigenvector (up to an overall phase) with eigenvalue one of the $n$ operators $K_j$.
The \emph{Clifford group} ${\cal C}_1$ on one qubit is the group of all $2\times 2$ unitary
operators which map $\sigma_u$ to $\alpha_u\sigma_{\pi(u)}$ under conjugation, where $u=x,y,z$, for
some $\alpha_u=\pm 1$ and  some permutation $\pi$ of $\{x,y,z\}$. The $n$-qubit local Clifford
group ${\cal C}_n^l$ is the $n$-fold tensor product of ${\cal C}_1$ with itself. The action of
local Clifford operations on graph states can be translated elegantly in terms of graph
transformations. In ref. \cite{localcliffgraph} we give such a translation and show that two states
$|G\rangle$, $|G'\rangle$ are LC-equivalent if and only if $G'$ can be obtained from $G$ by a
finite sequence of \emph{local complementations}. The local complement $g_i(G)$ of a graph $G$ at
one of its vertices $i\in V$ is defined by its adjacency matrix $g_i(\theta)$ as follows:
\begin{equation}\label{localcomp}g_i(\theta) = \theta + \theta_i\theta_i^T + \Lambda,\end{equation}
where $\theta$ is the adjacency matrix of $G$, $\theta_i$ is its $i$th column and $\Lambda$ is a
diagonal matrix such as to yield zeros on the diagonal of $g_i(\theta)$. Addition in
(\ref{localcomp}) is to be performed modulo two. In graph theoretical terminology, $g_i(G)$ is
obtained by replacing the subgraph of $G$ induced by the neighborhood of $i$ by its complement.


It is well known (see e.g. \cite{QCQI}) that the stabilizer formalism has an equivalent formulation
in terms of binary linear algebra. In this binary formulation, a stabilizer state 
on $n$-qubits corresponds to an $n$-dimensional self-dual linear subspace of $\Bbb{F}_2^{2n}$. Here
$\Bbb{F}_2$ is the finite field of two elements (0 and 1), where arithmetics are performed modulo
2. The self-duality of the subspace is with respect to a symplectic inner product $<\cdot, \cdot>$
defined by $< u, v>:= u^T P v$, where $u,v\in\Bbb{F}_2^{2n}$ and \[P = \left[
\begin{array}{cc} 0 & I \\ I& 0 \end{array} \right].\]
The binary stabilizer subspace is usually presented in terms of a full rank $2n\times n$ generator
matrix $S$, the columns of which form a basis of the subspace. This generator matrix satisfies
$S^TPS=0$ from the self-duality of the space. The entire binary stabilizer space, which we denote
by ${\cal C}_S$, is the column space of $S$.
It can easily be shown that a graph state with adjacency matrix $\theta$ has a generator matrix
\[ S =\left [ \begin{array}{c} \theta\\
I \end{array}\right].\] In the binary framework, local Clifford operations $U\in {\cal C}_n^l$
correspond to nonsingular $2n\times 2n$ binary matrices $Q$ of the block form
\[Q = \left [ \begin{array}{cc} A&B\\
C&D \end{array}\right],\] where the $n\times n$ blocks $A, B, C, D$ are diagonal
\cite{localcliffgraph}. We denote the diagonal entries of $A, B, C, D$, respectively, by
 $a_i$, $b_i$, $c_i$, $d_i$, respectively. The $n$ submatrices \[Q^{(i)}:=\left [
\begin{array}{cc} a_i & b_i
\\ c_i& d_i\end{array} \right ]\] correspond to the tensor factors of $U$. It follows
that each of the matrices $Q^{(i)}$ is invertible. Equivalently, the determinants $a_id_i+b_ic_i$
are equal to one. We denote the group of all such $Q$ by $C^l_n$.

We are now in a position to state the algorithm. Let $|G\rangle$, $|G'\rangle$ be two states with
adjacency matrices $\theta$, $\theta'$, respectively, and generator matrices \[S := \left[
\begin{array}{c} \theta\\ I
\end{array}\right],\ S' := \left[ \begin{array}{c} \theta'\\ I
\end{array}\right],\] respectively. Then $|G\rangle$ and $|G'\rangle$  are LC-equivalent if
and only if there exists $Q\in C^l_n$ such that ${\cal C}_{QS} = {\cal C}_{S'}$.
Equivalently, this occurs iff there exists an invertible $n\times n$ matrix $R$ over $\Bbb{F}_2$
such that
\begin{equation}\label{QSR1}
QSR=S'.
\end{equation}
If $\theta$ and $\theta'$ are given, (\ref{QSR1}) is a matrix equation in the unknowns $Q$ and $R$.
Note that we can get rid of the unknown $R$, as (\ref{QSR1}) is equivalent to
\begin{equation}\label{stelsel2}
S^T Q^T P S' = 0.
\end{equation}
Indeed, (\ref{stelsel2}) expresses that $u^T P v = 0$ for every $u\in {\cal C}_{QS}$ and $v\in
{\cal C}_{S'}$, which implies that ${\cal C}_{QS}$ and ${\cal C}_{S'}$ are each other's symplectic
orthogonal complement. These spaces must therefore be equal, as any $n$-dimensional binary
stabilizer space is its own symplectic dual, and (\ref{QSR1}) is obtained. More explicitly,
(\ref{stelsel2}) is the system of $n^2$ linear equations
\begin{equation}\label{alg}
\left(\sum_{i=1}^n \theta_{ij}\theta'_{ik}c_i\right)  + \theta_{jk}a_k + \theta'_{jk}d_j +
\delta_{jk}b_j = 0,
\end{equation}
for all $j,\ k=1, \dots, n$, where the $4n$ unknowns $a_i, b_i, c_i, d_i$ must satisfy the
quadratic constraints \begin{equation}\label{constraints}a_id_i+ b_i c_i =1.\end{equation}

The set ${\cal V}$ of solutions  to the linear equations (\ref{alg}), with disregard of the
constraints, is a linear subspace of $\Bbb{F}_2^{4n}$. A basis $B=\{b_1,\dots, b_d\}$ of ${\cal V}$
can be calculated efficiently in ${\cal O }(n^4)$ time by standard Gauss elimination over
$\Bbb{F}_2$. Then we can search the space ${\cal V}$ for a vector which satisfies the constraints
(\ref{constraints}). As (\ref{alg}) is for large $n$ a highly overdetermined system of equations,
the space ${\cal V}$ is typically low-dimensional. Therefore, in the majority of cases this method
gives a quick response. Nevertheless, in general one cannot exclude that the dimension of ${\cal
V}$ is of order ${\cal O}(n)$ and therefore the overall complexity of this approach is
nonpolynomial. However, it was shown in  \cite{alg_bouchet} that it is sufficient to enumerate a
specified subset ${\cal V'}\subseteq {\cal V}$ with $|{\cal V'}|={\cal O }(n^2)$ in order to find a
solution which satisfies the constraints, if such a solution exists. Indeed, the following lemma
holds:

\vspace{1mm}

\textbf{Lemma 1} \cite{alg_bouchet} {\it If dim$({\cal V})>4$, then the system
(\ref{alg})-(\ref{constraints}) of linear equations plus constraints has a solution  if and only if
the set \[{\cal V'}:= \left\{ b +b' \ |\ b, b'\in B \right\}\subseteq {\cal V}\] contains a vector
which satisfies the constraints.}

The proof of lemma 1 is involved and makes extensive use of local graph complementation. The reader
is referred to ref. \cite{alg_bouchet} for more details. Lemma 1 shows that, if a solution to
(\ref{alg})-(\ref{constraints}) exists, this solution can be found by enumerating either all
$|{\cal V}|\leq 16$ elements of ${\cal V}$ if dim$({\cal V})\leq4$ or the ${\cal O}(n^2)$ elements
of ${\cal V'}$ if dim$({\cal V})>4$ and checking these vectors against the constraints
(\ref{constraints}). Hence, a polynomial time algorithm to check the solvability of
(\ref{alg})-(\ref{constraints}) is obtained. The overall complexity of the algorithm is ${\cal
O}(n^4)$. Note that, whenever LC-equivalence occurs, this algorithm provides an explicit local
unitary operator in the Clifford group which maps the one state to the other, as a solution
$(a_1,b_1,c_1,d_1,\ \dots,\ a_n,b_n,c_n,d_n)$ to (\ref{alg})-(\ref{constraints}) immediately yields
an operator $Q\in C^l$.

In conclusion, we have presented an algorithm of polynomial complexity which detects whether two
given graph states are equivalent under the local Clifford group. This algorithm leans heavily on a
former result of ours, which is a description of the action of local Clifford operations on graph
states in terms of local graph complementation. Whenever equivalence of two graph states is
recognized, the algorithm provides an explicit local Clifford operator which maps the one state to
the other. Moreover, together with existing algorithms, this result yields an efficient algorithm
which recognizes local Clifford equivalence of all stabilizer states.

\begin{acknowledgments}
MVDN thanks M. Hein, for interesting discussions concerning local equivalence of stabilizer states,
and G. Royle, for pointing out the work of Bouchet. Dr. Bart De Moor is a full professor at the
Katholieke Universiteit Leuven, Belgium. Research supported by Research Council KUL: GOA-Mefisto
666, GOA-Ambiorics, several PhD/postdoc and fellow grants; Flemish Government: - FWO: PhD/postdoc
grants, projects, G.0240.99 (multilinear algebra), G.0407.02 (support vector machines), G.0197.02
(power islands), G.0141.03 (Identification and cryptography), G.0491.03 (control for intensive care
glycemia), G.0120.03 (QIT), G.0452.04 (QC), G.0499.04 (robust SVM), research communities (ICCoS,
ANMMM, MLDM); -   AWI: Bil. Int. Collaboration Hungary/ Poland; - IWT: PhD Grants, GBOU (McKnow)
Belgian Federal Government: Belgian Federal Science Policy Office: IUAP V-22 (Dynamical Systems and
Control: Computation, Identification and Modelling, 2002-2006), PODO-II (CP/01/40: TMS and
Sustainibility); EU: FP5-Quprodis;  ERNSI; Eureka 2063-IMPACT; Eureka 2419-FliTE; Contract
Research/agreements: ISMC/IPCOS, Data4s, TML, Elia, LMS, IPCOS, Mastercard; QUIPROCONE; QUPRODIS.

\end{acknowledgments}

\bibliographystyle{unsrt}
\bibliography{alg_codes}

\begin{thebibliography}{10}

\bibitem{Gott}
D.~Gottesman.
\newblock {\em Stabilizer codes and quantum error correction}.
\newblock PhD thesis, Caltech, 1997.
\newblock quant-ph/9705052.

\bibitem{graphbriegel}
W.~D\"ur, H.~Aschauer, and H.J. Briegel.
\newblock Multiparticle entanglement purification for graph states.
\newblock {\em Phys. Rev. Lett.}, 91:107903, 2003.
\newblock quant-ph/0303087.

\bibitem{1wayQC}
R.~Raussendorf, D.E. Browne, and H.J. Briegel.
\newblock Measurement-based quantum computation with cluster states.
\newblock {\em Phys. Rev. A}, 68:022312, 2003.
\newblock quant-ph/0301052.

\bibitem{entgraphstate}
M.~Hein, J.~Eisert, and H.J. Briegel.
\newblock Multi-party entanglement in graph states.
\newblock quant-ph/0307130.

\bibitem{localcliffgraph}
M.~Van~den Nest, J.~Dehaene, and B.~De~moor.
\newblock Graphical description of the action of local clifford operations on
  graph states.
\newblock {\em Phys. Rev. A}, 69:022316, 2004.
\newblock quant-ph/0308151.

\bibitem{Bouchet}
A.~Bouchet.
\newblock Recognizing locally equivalent graphs.
\newblock {\em Discrete Math.}, 114:75--86, 1993.

\bibitem{alg_bouchet}
A.~Bouchet.
\newblock An efficient algorithm to recognize locally equivalent graphs.
\newblock {\em Combinatorica}, 11(4):315 --329, 1991.

\bibitem{stabgraphcode}
D.~Schlingemann.
\newblock Stabilizer codes can be realized as graph codes.
\newblock quant-ph/0111080.

\bibitem{graph_grassl}
M.~Grassl, A.~Klappenecker, and M.~Roetteler.
\newblock Graphs, quadratic forms and quantum codes.
\newblock In {\em IEEE international symposium on information theory},
  Lausanne, 2001.

\bibitem{QCQI}
I.~Chuang and M.~Nielsen.
\newblock {\em Quantum computation and quantum information}.
\newblock Cambridge University press, 2000.

\end{thebibliography}

\end{document}